\documentclass[10pt,letterpaper]{article}
\usepackage{color}
\usepackage{graphicx}
\usepackage{dcolumn}
\usepackage{bm}
\usepackage{upgreek}
\usepackage{opex3}
\usepackage{cite}
\usepackage{amsmath}
\usepackage{url}
\begin{document}

\title{Deterministic generation of bright single resonance fluorescence photons from a Purcell-enhanced quantum dot-micropillar system}

\author{Sebastian Unsleber,$^{1,2}$ Christian Schneider,$^{1,2,*}$ Sebastian Maier,$^1$ Yu-Ming He,$^{1,3}$ Stefan Gerhardt,$^1$ Chao-Yang Lu,$^3$ Jian-Wei Pan,$^3$ Martin Kamp$^1$ and Sven H\"ofling$^{1,3,4}$}


\address{
$^1$Technische Physik and Wilhelm Conrad R\"ontgen Research Center for Complex Material Systems, Physikalisches Institut,
Universit\"at W\"urzburg, Am Hubland, D-97074 W\"urzburg, Germany\\
$^2$These authors contributed equally\\
$^4$SUPA, School of Physics and Astronomy, University of St Andrews, St Andrews, KY16 9SS, United Kingdom\\
$^3$Hefei National Laboratory for Physical Sciences at the Microscale and Department of Modern Physics,
$\&$ CAS Center for Excellence and Synergetic Innovation Center in Quantum Information and Quantum Physics,
University of Science and Technology of China, Hefei, Anhui 230026, China
}
\email{$^*$christian.schneider@physik.uni-wuerzburg.de}




\begin{abstract}
We report on the observation of bright emission of single photons under pulsed resonance fluorescence conditions from a single quantum dot (QD) in a micropillar cavity. The brightness of the QD fluorescence is greatly enhanced via the coupling to the fundamental mode of a micropillar, allowing us to determine a single photon extraction efficiency of $(20.7\pm0.8)~\%$  per linear polarization basis. This yields an overall extraction efficiency of $(41.4\pm1.5)~\%$  in our device.  We observe the first Rabi-oscillation in a weakly coupled quantum dot-micropillar system under coherent pulsed optical excitation, which enables us to deterministically populate the excited QD state. In this configuration, we probe the single photon statistics of the device yielding $g^{(2)}(0)=0.072\pm0.011$ at a QD-cavity detuning of $75~\mu$eV.
\end{abstract}






\section{Introduction}
Bright, efficient sources of single photons are key elements for quantum optics applications, including
quantum networks~\cite{Pan2012}, linear 
optical quantum computing~\cite{Kok2007,OBrien2007} and quantum teleportation~\cite{Nilsson-NatPhot13, Gao-NatCom13}. They furthermore play an important role in various boson sampling schemes, which can be exploited for photonic quantum emulation \cite{Aaronson2011,Spring2013}. 
Cold atoms, single ions, isolated molecules, optically active defects in diamonds, silicon carbide and layered materials, among others, have all been identified as attractive sources of non-classical light \cite{McKeever2004,Diedrich1987,Basche1992,Brouri00,Kurtstiefer2000,Castelletto2014,He2015,Tonndorf15}. 
In solid state, so far the most promising candidates for deterministic single-photon emission appear to be semiconductor quantum dots (QDs), because of their near-unity quantum efficiency and the possibility to achieve high interference visibilities~\cite{Michler2000,Santori2002,Gold2014}. Further advantages are the possibility to address them electrically~\cite{Yuan2002, Heindel2010, Ellis2008} and to integrate them in complex photonic environments and architectures, such as on-chip quantum optical networks~\cite{yao09,Hoang12}. 
Embedded in bulk semiconductor, however, QD-based single photon sources suffer from poor photon extraction efficiencies (similar to other solid state approaches), since only a minor 
fraction of the photons can leave the high refractive index material. 
This problem can be mitigated by integrating QDs into optical microcavities~\cite{Heindel2010,gazzano13,Gerard1998,Maier2014} 
or photonic waveguides~\cite{Claudon2010a,Heinrich2010,Reimer2012,arcari14}, where overall extraction efficiencies up to $\approx0.8$ photons per pulse have been demonstrated. The resonant
cavity approach furthermore enhances the spontaneous emission rate of single QDs via cavity quantum
electrodynamics (cQED), and as such can lead to significantly increased operation frequencies and furthermore is a viable tool to improve the indistinguishability of emitted photons \cite{Santori2002,Unsleber2015}. Alternatively, cavity stimulated Raman emission from single
dots has been explored in this context \cite{Sweeney2014}.

Carrier injection into single QDs is feasible via non-resonant optical and electrical injection. 
However, this non-resonant excitation introduces a time jitter on the excitation process. It was early recognized, that resonant driving of QDs can be used to deterministically populate the excited state of an exciton \cite{Zrenner2002} and even the biexciton \cite{Jayakumar2013} of a single QD, which is hard to achieve in non-resonant pumping schemes \cite{predojevic2014efficiency}. Furthermore, due to the reduction of time uncertainties in the population dynamics of the QD and the reduction of the carrier bath in the environment of the QD, resonant techniques promise to be superior for the generation of indistinguishable photon wave packets. Recently, we have demonstrated near unity indistinguishability from a resonantly driven single QD embedded in a planar microcavity with a low Q factor \cite{He2013}, which underlines the power of this approach.
Thus far, the generation of bright, on-demand resonance fluorescence photons has only been explored in planar structures, due to the experimental complication of suppressing the pump laser with high ratios. Microcavity assisted experiments have been carried out, however under weak pumping conditions in the regime of photon blockade \cite{Faraon2008,Reinhard2012}. 
Here, we demonstrate that polarization filtering is sufficiently effective to suppress the pump laser by approximately a factor of $10^7$ when scattered from a AlAs/GaAs micropillar cavity with a diameter of $4~\mu$m. We present results on the coherent control and deterministic injection of a QD exciton in the micropillar with a quality factor (Q-factor) of $5950$ and a Purcell factor of $F_P=3.0\pm0.6$. This allows us to detect single photons on demand from the resonantly driven system, with an overall extraction efficiency of the device up to $\eta_{ext}=(41.4\pm1.5)\%$ characterized by a low two photon probability of $g^{(2)}(0)=0.072\pm0.011$. Furthermore, we study the dynamics of the coupled system via measuring the dependence of the Rabi-oscillations for different QD-cavity and QD-laser detunings. \\

\section{Experimental details}
\begin{figure}[htbp]
\centering
\fbox{\includegraphics[width=\linewidth]{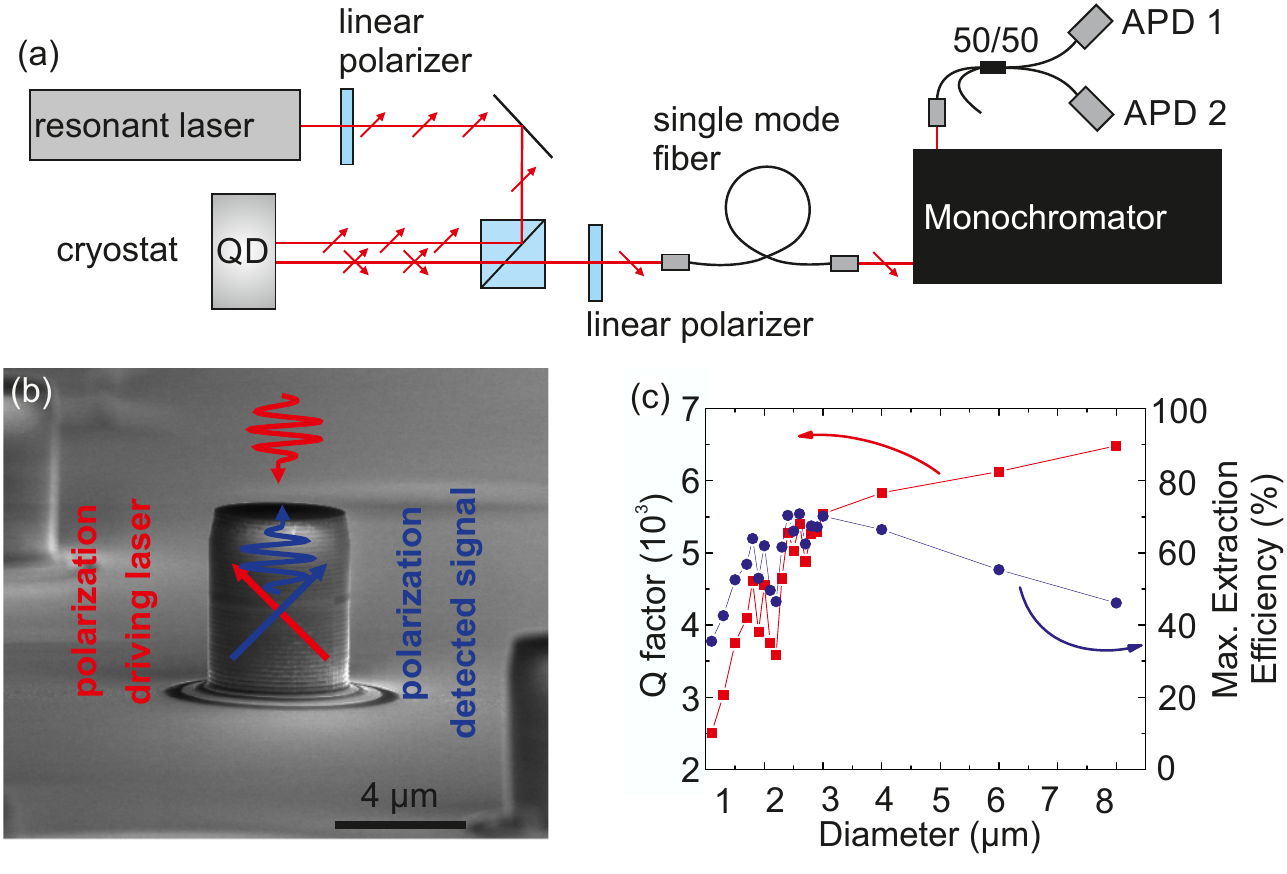}}
\caption{(a) Schematic drawing of the experimental setup. The resonant excitation is implemented by suppressing the laser in a cross-polarization configuration. (b) SEM picture of a micropillar with 4 $\mu$m diameter. (c) Q-factor and calculated maximal achievable extraction efficiencies versus the pillar diameter.}
\label{Fig1}
\end{figure}

Figure \ref{Fig1}(a) shows a schematic drawing of our experimental setup. The resonant, linearly polarized Ti:Sapphire laser (repetition rate $82$ MHz, pulse length $\tau\approx1.3$ ps) is coupled into the beam path via a $92/8$ pellicle beam splitter. The micropillar sample is mounted on the coldfinger of a liquid Helium flow cryostat and the emitted light from the sample is collected with a microscope objective (NA$=0.40$) and coupled into a single mode fiber. A second linear polarizer in front of the fiber coupler is orientated perpendicular to the laser polarization and selects the detected polarization axis of the QD signal. Via carefully adjusting both polarizers we can achieve a suppression of the scattered laser light by $\approx 10^7$. We can test the single photon properties of the emitted QD-photons via coupling the spectrally narrow filtered photons into a fiber based Hanbury Brown and Twiss (HBT) setup. The emitted photons are split at a $50:50$ beam splitter onto two Silicon-based avalanche photo diodes (APD's) with a timing resolution of $t_{Res}\approx400$ ps and the second order autocorrelation function is measured.
Figure \ref{Fig1}(b) shows a scanning electron microscope (SEM) image of a $4~\mu$m diameter micropillar, which consists of $25.5$ ($15$) $\lambda/4$-thick AlAs/GaAs mirror pairs which form the lower (upper) distributed Bragg reflector (DBR). The DBR stacks sandwich a $\lambda$-thick GaAs cavity with a layer of In(Ga)As-QDs as active medium. After growing the planar sample via means of molecular beam epitaxy, micropillars with varying diameters ranging from $1~\mu$m to $8~\mu$m were defined via electron beam lithography and transferred into the sample by electron-cyclotron-resonance reactive-ion-etching.
We first determined the quality factor $Q=E/\Delta E$ of our micropillar sample via standard, non-resonant $ß\mu PL$ measurements of the fundamental optical mode. The diameter dependent Q-factors are shown in Fig. \ref{Fig1}(c). For decreasing pillar diameters from $8~\mu$m to $1~\mu$m, we observe a characteristic decrease of the Q-factor as a result of increasing sidewall losses and increasing mode mismatch \cite{Reitzenstein2007}. The maximum extraction efficiency can be estimated via \cite{Barnes2002}
\begin{equation}
\eta_{ext}=\frac{Q_{Pillar}}{Q_{2D}}\text{x}\frac{F_{P,max}}{\gamma+F_{P,max}}
\label{equ:efficiency}
\end{equation}
with $Q_{Pillar}$ and $Q_{2D}$ being the Q-factors of the etched pillar and the planar microcavity, $F_{P,max}=3Q(\lambda_C/n)^3/(4\pi^2V_M)$ ($\lambda_C$: resonance wavelength; $n$: refractive index; $V_M$: mode volume) being the maximum Purcell enhancement \cite{Gayral03} and $\gamma$ being the fraction of the emission which is emitted into leaky modes. Note, that for a micropillar cavity, one can safely assume  $\gamma\approx1$ \cite{Barnes2002}.
In order to estimate the mode volume of the micropillar, we have extrapolated the values reported in  \cite{Boeckler2008} with the micropillar area. 
Using our experimentally measured Q-factors, we can assess the extraction efficiencies of our micropillar sample via equation \ref{equ:efficiency}. The result is shown in Fig. \ref{Fig1}(c). The calculation shows that we can expect extraction efficiencies above $60~\%$ for pillar diameters between $2~\mu$m and $4~\mu$m.  The Q-factor of the planar cavity was experimentally extracted to be $Q_{2D}=6670$.

\section{Experimental results and discussion}
\begin{figure}[!h]
\centering
\fbox{\includegraphics[width=\linewidth]{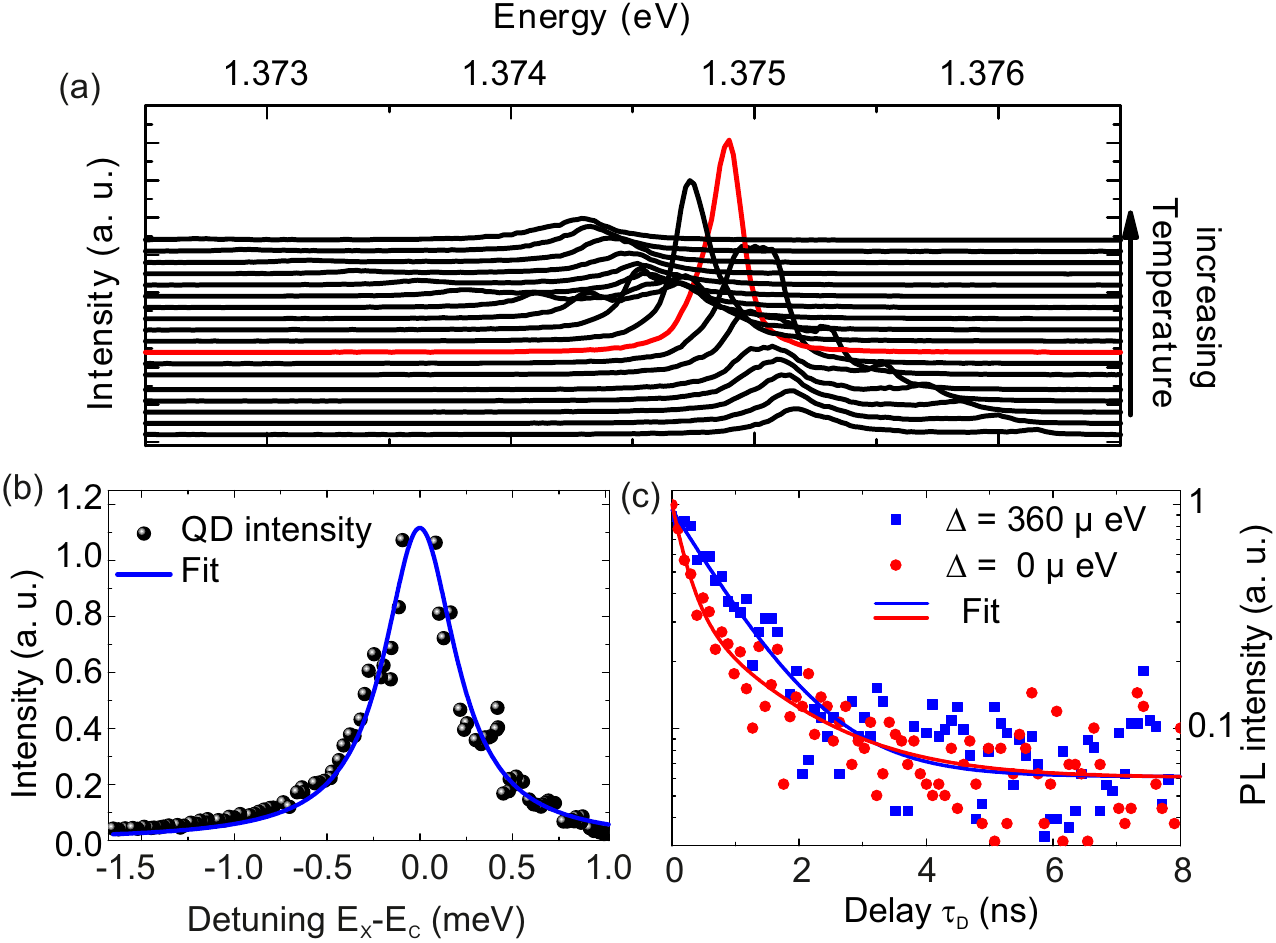}}
\caption{(a) Temperature dependent spectra of a $4~\mu$m pillar under above bandgap excitation. A strong enhancement on spectral resonance due to the Purcell effect is observed. (b) Integrated intensity of the QD emission under weak non-resonant pumping, significantly below the saturation of the QD. The fits suggest a Purcell enhancement of the system around 3. (c) Time resolved measurements on and off resonance for p-shell excitation reveal a Purcell factor of $F_P=3.0\pm0.6$.}
\label{Fig2}
\end{figure}
In the following, we present the study of a micropillar with a diameter of 4 $\mu$m (Q=5950, $\gamma_C=233~\mu$eV). Figure \ref{Fig2}(a) shows a series of spectra recorded under non-resonant excitation and varying temperature. We observe a clear signature of weak light matter coupling and a strong enhancement of the emitted QD intensity for the spectral resonance between the QD exciton and the fundamental optical mode due to the Purcell effect as the QD shifts through the fundamental mode with increasing temperature. In Fig. \ref{Fig2}(b), we analyzed the integrated intensity of the QD as a function of the exciton-cavity-detuning $\Delta$ under weak non-resonant pumping, significantly below the saturation of the QD. The intensity as a function of the exciton-cavity detuning can be expressed by $I_X(\Delta)\propto F_P/(F_P+1+\frac{\Delta^2}{\gamma_c^2})$ \cite{Munsch2009}. This expression yields $F_P=3.1\pm0.1$ when fitted to our experimental data, which is in very good agreement with the theoretical maximal Purcell enhancement in this pillar of $F_{P, theo. max.}=3.2$. Thus, we conclude, that the QD is located close to the center of the cavity. 
In order to further characterize the coupling strength between the QD and the optical mode, we carried out time-resolved $\mu PL$ measurements. To prevent carrier refilling, we excited the QD into a higher resonance, $12$ meV blue shifted with respect to the exciton.  Figure \ref{Fig2}(b) shows the time depended QD emission for spectral resonance between QD and fundamental mode (red squares) and for a spectral detuning of $\Delta=E_X-E_C=360~\mu$eV. In order to guarantee comparability between the two decay curves, we have subtracted the background which we could directly relate to the dark counts from the high resolution detectors ($t_{Res}\approx40$ ps). We fitted the resonant case  with a biexponential decay and the data off resonance with a mono-exponential curve. On resonance, the recorded signal is composed of emission from the probed emitter and the cavity background, which which is additionally illuminated by non-resonant spectator QDs. This leads to a slow decay of the non-resonantly illuminated cavity signal (approx. on the order of the QD lifetime $\approx1$ ns) \cite{Suff2009}. From the fits we obtain $T_1$-times of $T_1(\Delta=0~\mu\text{eV})=(221\pm40)$ ps and $T_1(\Delta=360~\mu\text{eV})=(890\pm66)$ ps. Since QD and cavity are detuned $\approx1.5*\gamma_C$, we assume the emission into leaky modes to dominate the off-resonant measurement, which therefore leads to a Purcell factor for this QD of $F_P=\frac{T_{off}}{T_{on}}-1=3.0\pm0.6$ \cite{Munsch2009}. To further confirm this result, we have recorded the decay time of 5 QDs after removing the top DBR on another sample piece from the same wafer, yielding an average decay time of $T_{bulk}=(852\pm177)$ ps which agrees very well with the off-resonant decay time. By applying a simple analytical model for the calculation of the lateral electric field in the pillar \cite{Gutbrod1998}, we can estimate a maximum misalignment of $\approx300$ nm from the cavity centre.
 \begin{figure}[bp]
\centering
\fbox{\includegraphics[width=\linewidth]{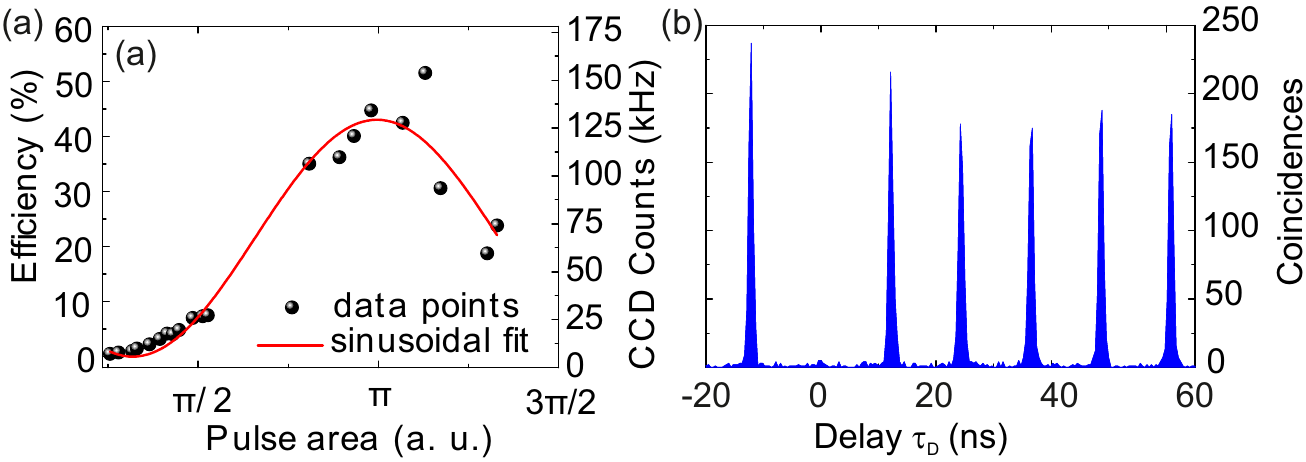}}
\caption{(a) Measured source efficiency and count rate on the monochromator CCD versus the pulse area of the driving laser field for a QD cavity detuning of $\Delta=75~\mu$eV. We extract a $g^{(2)}(0)$-corrected overall efficiency of $\eta=(41.4\pm1.5)~\%$. (b) 2$^{nd}$ order autocorrelation histogram for pulsed resonant excitation for a detuning of the QD of $\Delta=75~\mu$eV. We extract a $g^{(2)}$-value as low as $g^{(2)}(0)=0.072\pm0.011$.}
\label{Fig3}
\end{figure}
In the following, we address succeeding experiments which were carried out under strictly resonant, pulsed excitation conditions. First, we carefully calibrated the efficiency of our setup sketched in Fig. \ref{Fig1}(a), and extracted a value of $\eta_{Setup}=(0.36\pm0.02)\%$. This allows us to derive the extraction efficiency of the micropillar via measuring the intensity of the QD emission on the CCD of our spectrometer. The result of the power dependent measurement in the resonance fluorescence configuration are plotted in Fig. \ref{Fig3}(a). As expected, the intensity follows a sinusoidal behaviour, with a maximum for $\pi$ pulse excitation. We measure count rates on the order of 130 kHz on the CCD, which converts into a maximum extraction efficiency of $\eta_{lin}=(21.5\pm0.8)\%$. Note, that this value takes into account only photons from one linear polarization due to the filtering process. Accordingly, the overall extraction efficiency of the source acquires a value of $\eta_{Source}=(43.0\pm1.6)\%$. Due to an non-zero second order autocorrelation (see below), this value must be corrected by multi-photon pulses via the equation $\eta_{SPS}=\eta\sqrt{1-g^{(2)}(0)}$ leading to a single photon extraction efficiency of $\eta_{SPS}=(41.4\pm1.5)~\%$ (or $\eta_{polarized}=(20.7\pm0.8)~\%$ for one polarization). It is important to note, that for applications relying on indistinguishable photons, only linearly polarized photons can be used, irrespective of the excitation and filtering conditions. Therefore it is highly desirable to increase the source efficiency of linear polarized photons to the full device efficiency e. g. via coupling the QD to a linear polarized optical mode of the pillar \cite{Daraei2007}. 
The slight deviation from the theoretical maximum of $\approx~65~\%$, plotted in Fig. \ref{Fig1}(c), can be attributed to the spatial misalignment of the QD, as well as to the slight QD-cavity detuning of $\Delta=75~\mu$eV during the measurement.\\
Measurements of the second order autocorrelation function of the QD emission under $\pi$ pulse excitation were carried out by coupling the emitted photons into a fiber-based HBT setup. The measured coincidence histogram for a QD cavity detuning of $\Delta=75~\mu$eV is shown in Fig. \ref{Fig3}(b). For smaller detunings, we observed a cavity feeding effect of a spectator QD which is simultaneously driven by the pump laser, and consequently reduces the purity of the single photon emission. We extract the $g^{(2)}(0)$-value by dividing the area within a $12.2$ ns window of the central peak via the average area of the surrounding peaks yielding $g^{(2)}(0)=0.072\pm0.011$. These results unambiguously proof pure and bright single photon emission from a deterministically populated exciton state. We also like to mention, that we see a slight blinking effect which can often be observed for (quasi-) resonant excitation schemes and can be attributed to an "`off"'-state to which the QD changes after a certain time \cite{Santori2001}.\\
Furthermore, we study the influence of the QD cavity detuning on the resonance fluorescence properties of the QD. Figure \ref{Fig4} shows the integrated QD intensity for various detunings with respect to the pump laser and the microcavity resonance. The fits are sinusoidals with an additional exponential decay term which phenomenologically accounts for damping of the Rabi-oscillations due to coupling to longitudinal acoustic (LA) phonons or other sources of decoherence \cite{Ramsay2010a}.
\begin{figure}[h]
\centering
\fbox{\includegraphics[width=0.8\linewidth]{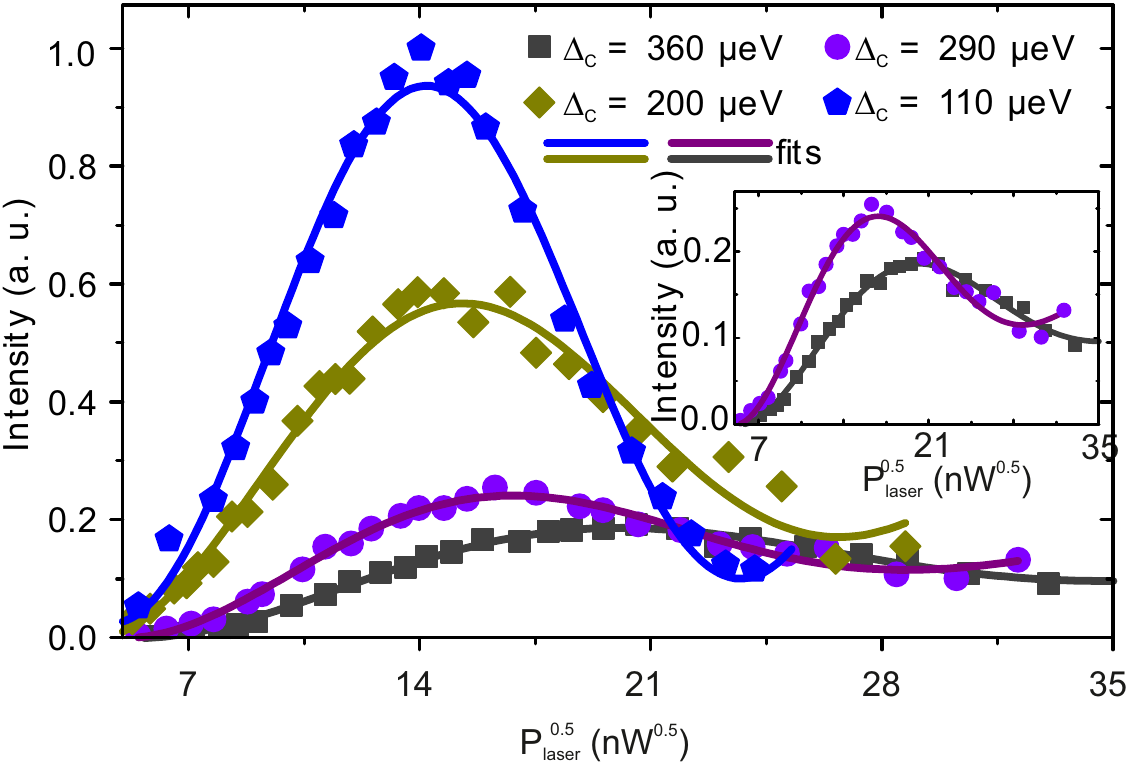}}
\caption{Rabi-oscillations recorded for various detunings between QD and cavity. The fits are sinusoidals with an additional exponential damping term. For an increased detuning we observe a drop in the maximum emitted intensity as well as an increasing damping of the Rabi-oscillations.}
\label{Fig4}
\end{figure}
For all conditions, we observe clear Rabi-oscillations which are expected for driving a two level system with a resonant pulsed laser field. As we increase the detuning, we observe three trends in the Rabi-oscillations. First, due to the lowered spectral overlap between QD and cavity, the extraction efficiency decreases and therefore the effective maximal integrated intensity for driving the system with an effective $\pi$-pulse is lowered by more than a factor of four. The second effect we observe is a shift of the $\pi$-pulse laser power. The QD is driven with a laser field, that is funnelled by the cavity mode. Therefore, if the spectral detuning between the QD and the microcavity resonance increases, the spectral overlap with the laser field gets smaller and the laser power which is needed to invert the two level system increases by a factor of $\approx 2$ in our experiment. The third effect we can observe is a damping of the Rabi-oscillations which gets stronger for larger detunings. This can have several possible origins: A constant dephasing can be introduced by defects in the vicinity of the QD or scattering with the sidewalls of the micropillar. Furthermore, the temperature dependence could origin from an increasing density of LA phonons towards higher temperatures (i. e. larger detunings) which leads to an increased damping of the Rabi oscillations \cite{Ramsay2010}. In addition, when detuning the laser field from the two level system, the QD pumping is increasingly  phonon assisted and therefore the oscillation amplitude also gets lowered \cite{Ardelt2014}.\\
\section{Conclusion}
In conclusion, we have demonstrated efficient, deterministic generation of single photons on demand in a QD-micropillar system via pulsed resonant pumping. Our device reveals an outcoupling efficiency of $\eta=(41.4\pm1.5)~\%$ combined with very low probabilities for multi-photon emission events characterized by $g^{(2)}(0)=0.072\pm0.011$.
In addition, we have shown, that a QD can couple to two different light fields at the same time. We observe a strong Purcell-enhancement of the spontaneous emission for minimizing the spectral overlap between QD and fundamental cavity mode which is a clear signature of weak light matter interaction. Secondly, we report on Rabi-oscillations from such a QD-cavity system, which is an unambiguous proof for the coupling of the QD to the driving laser field via the microcavity.
This experimental implementation combining deterministic generation of a QD exciton, very high photon extraction efficiencies, significant spontaneous emission enhancement and the possibility to achieve very high source coherence is a significant step towards advanced interference experiments based on single photons.


\section*{Funding Information}
 We acknowledge financial support by the State of Bavaria and the German Ministry of Education and Research (BMBF) within the projects Q.com-H and the Chist-era project SSQN. 
Y.-M. H. acknowledges support from the Sino-German (CSC-DAAD) Postdoc Scholarship Program.

\section*{Acknowledgments}

The authors would like to thank M. Emmerling for expert sample preparation. C.S. would like to thank Dara McCutcheon and Niels Gregersen for numerous discussions about combining brightness and coherence in a pillar single photon source.



\end{document}